\documentclass[12pt]{article}
\usepackage[dvips]{graphicx}
\usepackage{latexsym}
\newcommand{\beq}{\begin{equation}}
\newcommand{\eeq}{\end{equation}}
\newcommand{\Frac}[2]{\frac{\displaystyle #1}{\displaystyle #2}}

\newcommand{\Oa}{{\cal O} (\alpha_{s}^3)}
\newcommand{\Oaa}{{\cal O} (\alpha_{s}^2)}

\begin{document}
\thispagestyle{empty}
\begin{titlepage}
\begin{center}
\vspace*{3.5cm} 
\begin{Large}
{\bf On the massless contributions to the vacuum polarization of heavy
quarks
} \\[2.25cm]
\end{Large}
{ \sc J. Portol\'es} \ and { \sc P. D. Ruiz-Femen\'\i a}\\[0.5cm]
{\it Departament de F\'\i sica Te\`orica, IFIC, Universitat de Val\`encia -
CSIC\\
 Apt. Correus 22085, E-46071 Val\`encia, Spain }\\[2.5cm]

\begin{abstract}
\noindent
In Ref.~\cite{pg1} Groote and Pivovarov have given notice of a possible
fault in the use of sum rules involving two--point correlation
functions to extract information on heavy quark parameters, due to the
presence of massless contributions that invalidate the construction of
moments of the spectral densities. Here we show how to circumvent this
problem through a new definition of the moments, providing an infrared
safe and consistent procedure.

\end{abstract}
\end{center}
\vfill
\hspace*{1cm} PACS~: 11.55.Hx, 11.55.Fv, 13.65.+i, 12.38.Bx \\
\hspace*{1cm} Keywords~: Heavy quark sum rules, singularities of perturbative
amplitudes.
\eject
\end{titlepage}

\pagenumbering{arabic}

\section{Introduction}
\hspace*{0.5cm}The vacuum polarization function suitable for extracting
fundamental information of heavy quark-antiquark systems is built from the
electromagnetic current $j^\mu (x) = e_Q \, 
\overline{Q}(x)\, \gamma^\mu \, Q(x)$ of the 
heavy quark $Q$ of mass $M$ and $e_Q$ electric charge~: 
\begin{equation}
\Pi_{\mu \nu}(q) =  i \int d^4x \, e^{iqx}\,
\langle 0| T \,j_{\mu}(x) \,
j_{\nu}^{\dagger}(0) \,|0\rangle \,=\, 
\left( -g_{\mu\nu}q^2+q_{\mu}q_{\nu} \right)\Pi(q^2)
\,.
\label{eq:vacumm}
\end{equation} 
As it is well known two--point functions are analytic except for 
singularities at simple poles or branch cuts, the latter being originated
by normal thresholds of production of internal on--shell states.
Implicitly assuming that the absorptive part of $\Pi(q^2)$ starts at the
massive two--particle threshold $q^2 = 4 M^2$, vanishing below this point,
the correlator satisfies the once--subtracted dispersion relation 
\cite{dera} \footnote{Sometimes de Adler function defined as
$\partial \Pi(q^2) / \partial \ln q^2$, to get rid of the subtraction
constant, is used.}~:
\begin{equation}
\widehat{\Pi}(q^2) \, \doteq \, 
\Pi(q^2) \, - \, \Pi(0) \,  = \, \frac{q^2}{\pi}\int^{\infty}_{4M^2}
\Frac{ds}{s} \,\,
\frac{\mbox{Im}\,\Pi(s)}{s-q^2-i\epsilon}
\; \; .
\label{eq:disp-rel}
\end{equation}
This dispersion relation has been extensively used to determine heavy
quark parameters within the method of sum rules because it allows to
relate experimental input, on the right--hand side, with theoretical
perturbative evaluations on the left--hand side \cite{rev}. Indeed
$\mbox{Im} \, \Pi(q^2)$ refers to the total cross section
of heavy quark production $\sigma (e^+ e^- \rightarrow Q \overline{Q})$.
In practice, the 
spectral density $\mbox{Im} \, \Pi (s)$ is poorly known 
experimentally at very high energies and, in addition, we are interested
in the very low energy domain because it is more sensitive to the
heavy quark mass. Therefore one uses derivatives of the
vacuum polarization at the origin, called moments, to be responsive
to the threshold region~:
\begin{equation} 
{\cal M}_n=\frac{1}{n!}\left(\frac{d}{dq^2}\right)^n\Pi(q^2)\arrowvert_{q^2=0}
\; \; \; .
\label{eq:def_moments}
\end{equation} 
\par
Until present the evaluation of the perturbative two--point correlation
function $\Pi (q^2)$ has only been carried out completely, with massive
quarks, up to ${\cal O} (\alpha_s^2)$ \cite{cher}
and the procedure above has been termed consistent and effective in 
its task because the first branch point is set at the massive two--particle
threshold.
However in Ref.~\cite{pg1} Groote and Pivovarov have pointed out that at 
$\Oa$ there is a contribution to the
correlator which contains a three--gluon massless intermediate state 
(see Fig.~\ref{fig:3gluon}(a)).
Its absorptive part starts at zero energy and, therefore, 
Eq.~(\ref{eq:disp-rel}) is no longer correct. Moreover those authors have
also warned about the fact that, at this perturbative order, the massless
intermediate state invalidates the definition of the moments ${\cal M}_n$
for $n \ge 4$ because they become singular. In Ref.~\cite{pg2} an
infrared safe redefinition of the moments, to cure the latter problem,
has been provided; it consists in evaluating the moments at an 
Euclidean point $q^2 = - s_E$, $s_E > 0$, thus avoiding the singular behaviour.
Nevertheless the fault in Eq.~(\ref{eq:disp-rel}) due to the massless
threshold still represents a problem because even if, as we will
justify later on, we substitute this
dispersion relation by 
\begin{equation} 
\widehat{\Pi}(q^2) \;  = \; \Frac{q^2}{\pi}\int^{\infty}_{4M^2}
\Frac{ds}{s} \, \, 
\frac{\mbox{Im}\,\Pi_{ Q\overline{Q}}(s)}{s-q^2-i\epsilon}\; + \; 
\Frac{q^2}{\pi}\int^{\infty}_{0} \Frac{ds}{s} \,\,
\frac{\mbox{Im}\,\Pi_{3g}(s)}{s-q^2-i\epsilon}
\; \; \; ,
\label{eq:fulldisp-rel}
\end{equation}
(where the notation is self-explicative), 
the spectral
function $\mbox{Im} \, \Pi_{3g}(s)$ associated to the cut in 
Fig.~\ref{fig:3gluon}(a) would hardly be implemented phenomenologically 
as gluons hadronize to both heavy and light quark pairs. 
Perturbatively the three--gluon cut would contribute to 
$Q \overline{Q}$ production, i.e. to 
$\mbox{Im} \, \Pi_{Q \overline{Q}}(q^2)$,
but at higher order in $\alpha_s$. Therefore if we attach to an
$\Oa$ sum rule analysis, that contribution should be extracted from the
perturbative $\Pi(q^2)$ evaluation.
In this note we provide a bypass to recover the balance between the
right-hand and left-hand parts of Eq.~(\ref{eq:fulldisp-rel}).

\section{Moments and the massless cut}
\begin{figure}[tb]
\begin{center}
\hspace*{-0.5cm}
\includegraphics[angle=0,width=0.7\textwidth]{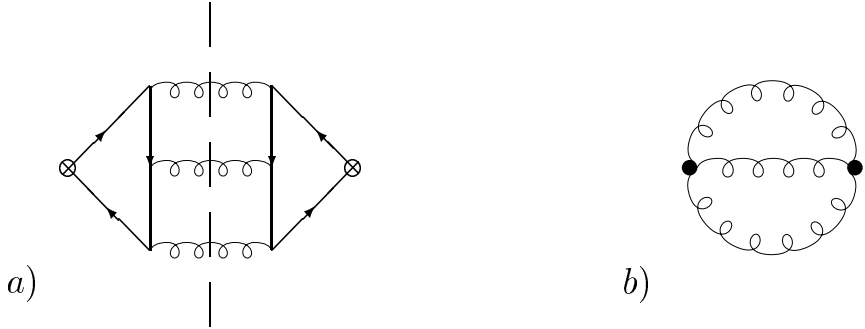}
\end{center}
\caption[]{\label{fig:3gluon} \it (a) $\Oa$ diagram contributing to the
vacuum polarization function of the heavy quark current (the vertical 
dashed line indicates the massless cut). (b) \lq \lq Effective" diagram
obtained by integrating out the fermion loops. It also has the topological
structure of the \lq \lq reduced" diagram that determines the massless
cut singularity.}
\end{figure}

\hspace*{0.5cm}
The perturbative contribution given by the diagram in 
Fig.~\ref{fig:3gluon}(a)
has been calculated at small $q^2$ ($q^2 \ll M^2$) in 
Ref.~\cite{pg1}. In this limit the quark triangle loop can be integrated
out and it ends up in the diagram in Fig.~\ref{fig:3gluon}(b) generated
by an induced
effective current describing the interaction of the vector current
with three gluons \footnote{The permutations of the three gluons in 
Fig.~\ref{fig:3gluon}(a) are already included in the definition
of the effective current.},
\begin{equation} 
J^{\mu}\,=\, - \Frac{\pi}{180 M^4} \, \left( \Frac{\alpha_s}{\pi} 
\right)^{\frac{3}{2}} \, \left(5\,
\partial_{\nu}{\cal
O}_{1}^{\mu\nu}\,+\,14\,\partial_{\nu}{\cal O}_{2}^{\mu\nu}\right)
\,,
\label{eq:effec_current}
\end{equation}

\noindent
with
\begin{eqnarray} 
{\cal O}_{1}^{\mu\nu} \,  & = & \, d_{abc}\,G^{\mu\nu}_a G^{\alpha\beta}_b 
G_{\alpha\beta}^c \; \; , \\ \nonumber
{\cal O}_{2}^{\mu\nu} \, & = & \, d_{abc}\,G^{\mu\alpha}_a G_{\alpha\beta}^b 
G^{\beta\nu}_c \; \; ,
\label{eq:operators} 
\end{eqnarray}  
where $G^{\mu\nu}_a$ is the gluon strength field tensor. The effective 
current in the QED case 
($G^{\mu\nu}_a\to F^{\mu\nu}, \alpha_s \to \alpha_{em}, d_{abc}\to 1$) can be 
easily identified from the
lowest order Euler-Heisenberg Lagrangian (see Ref.~\cite{pg2}).  
\par
The correlator of the induced current (\ref{eq:effec_current}) is then
evaluated in the configuration space giving~:
\begin{equation}
\langle 0| T \,J_{\mu}(x) \
J_{\nu}^{\dagger}(0)\, |0\rangle\,=\,-\frac{34}{2025\pi^4 M^8}
\left(\frac{\alpha_s}{\pi}\right)^3 d_{abc}d_{abc}\,
\left(\partial_{\mu}\partial_{\nu}-g_{\mu\nu}\partial^2\right)
\frac{1}{x^{12}} \, .
\label{eq:correlator}
\end{equation}
In momentum space we need to perform the Fourier transform of
Eq.~(\ref{eq:correlator}). Following the
differential regularization procedure \cite{differ}, which works
directly in configuration space, the result for the vacuum polarization
contribution of the diagram in Fig.~\ref{fig:3gluon}(b) at small $q^2$ reads
\begin{equation} 
 \Pi_{\mu\nu}(q)\; = \; \frac{17}{2916000 \pi^2}\,d_{abc}d_{abc}
\left(\frac{\alpha_s}{\pi}\right)^3
(q_{\mu}q_{\nu}-q^2g_{\mu\nu})\left(\frac{q^2}{4M^2}\right)^4
\ln \left(\frac{\mu^2}{-q^2}\right)\,
+{\cal O}\Big[ \Big(\frac{q^2}{M^2}\Big)^5 \Big]\,,
\label{eq:3gluon_polarization}
\end{equation}
with $\mu$ the renormalization point in this scheme,
and $d_{abc}d_{abc}=40/3$. 
\par
As noticed by Groote and Pivovarov \cite{pg1}, moments associated to the
diagram in Fig.~\ref{fig:3gluon}(b) are not defined if $n\ge 4$. 
Indeed differentiating 
Eq.~(\ref{eq:3gluon_polarization}) four times, at $q^2\approx 0$, we get:
\begin{equation}
\frac{1}{4!}\left(\frac{d}{dq^2}\right)^4\Pi(q^2)\arrowvert_{q^2\approx 0}=
\frac{17}{218700\pi^2}\left(\frac{\alpha_s}{\pi}\right)^3
\left(\frac{1}{4M^2}\right)^4
\left[\ln \left(\frac{\mu^2}{-q^2}\right)-\frac{25}{12}\right]
+{\cal O}\Big[ \frac{q^2}{M^{10}} \Big]
\; \; ,
\label{eq:moment4}
\end{equation}
whose real part clearly diverges if we set $q^2=0$. Larger $n$ moments are also
infrared divergent, and so the authors of Ref.~\cite{pg1} conclude
that the standard sum rule analysis must limit the accuracy of theoretical
calculations for the $n\ge 4$ moments to the ${\cal O}(\alpha_s^2)$ order of
perturbation theory. 
\par
One obvious way out of this infrared problem is to avoid the $q^2=0$ point.
As it has been discussed in Ref.~\cite{pg2}, this solution is rather 
ill--conditioned from the phenomenological side though. Moreover we notice
(as commented earlier) that it is not possible to implement, from the 
available experimental information, the second term in the right--hand side
of Eq.~(\ref{eq:fulldisp-rel}). However, if one does not insist in using full 
vacuum polarization for the sum rule analysis there is a way to 
overcome this infrared problem.

\section{Infrared safe definition of the moments}
\hspace*{0.5cm} The study
of analytic properties of perturbation theory amplitudes shows that their
singularities are isolated and, therefore, we can discuss each singularity
of a perturbative amplitude by itself \cite{oldies}. As a consequence, 
$\Pi (q^2)$ in 
Eq.~(\ref{eq:vacumm}) satisfies a dispersion relation from Cauchy's 
theorem \footnote{This expression also gives the residue $R_i$ of a pole at 
$s = s_i$ if we interpret the discontinuity as 
$\mbox{Im} \Pi_i \, = \, \pi R_i \delta(s-s_i)$. However we do not consider
the existence of $Q \overline{Q}$ Coulomb bound states, as it is not 
relevant for our discussion.}~:
\begin{equation}
\widehat{\Pi} (q^2) \, = \, \sum_n \, \Frac{q^2}{2 \pi i} \, 
\int_{s_n}^{\infty} \, 
\Frac{ds}{s} \, \Frac{\left[  \, \Pi(s) \, \right]_n}{s \, - \, 
q^2 \, - \, i\epsilon} \; \; \; .
\label{eq:piq2}
\end{equation}
Here
$\left[ \Pi(s) \right]_n$ provides the discontinuity across a 
branch cut starting at the branch point $s_n$. 
\par
In the perturbative calculation,
every discontinuity function $\left[ \Pi(s) \right]_n$
can be associated to a \lq \lq reduced" Feynman
diagram obtained by contracting internal off--shell propagators to a point
and leaving internal on--shell lines untouched.
Its contribution is written down following the Cutkosky rules for the
graph. The reduced diagram 
corresponding to the massless cut in Fig.~\ref{fig:3gluon}(a) has the
topological structure of the part (b) of that Figure. Let us emphasize though
that our following discussion is not grounded on the $q^2 \ll M^2$ regime
where the fermion loops have been integrated out~:
the reduced diagram is just a symbol that specifies a singularity, and the
black dots in Fig.~\ref{fig:3gluon}(b) keep all the analytical structure of
the fermion loops.
\par
In a general diagram the discontinuity across a specified cut needs not to
be a pure real function in the physical region, only the sum of all cuts 
in a given channel gives
the total imaginary part. Hence the separation between the imaginary parts
coming from different final states, as performed
in Eq.~(\ref{eq:fulldisp-rel}) for the vacuum polarization, does not seem
to come directly from the
Cutkosky rules.
Nevertheless in the heavy quark correlator the discontinuity across the 
three--gluon cut
gives a contribution to the spectral function that is unequivocally 
real~:
\begin{equation}
\Frac{1}{2i} \, \left[ \, \Pi(s) \, \right]_{3g} \, = \,  \mbox{Im} \, 
\Pi_{3g} (s)
 \; = \; - \, \Frac{1}{6s}\int \, d R_{3g} \,
\langle \, 0 \, | \, j^{\mu} \, | \, 3 \, g \, \rangle \, \langle \, 3 \, 
g \, | \, j_{\mu}^{ \dagger} \, | \, 0 \, \rangle  \; \; ,
\label{eq:unita}
\end{equation}
from which the dispersive part can be evaluated independently of the
$Q \overline{Q}$ cuts \footnote{The integration
in Eq.~(\ref{eq:unita}) extends to the available three--gluon phase
space.}. 
Accordingly we conclude that we can identify and isolate the troublesome
massless cut contribution to the two--point function. Indeed 
Eqs.~(\ref{eq:piq2}) and (\ref{eq:unita}) justify our previous
Eq.~(\ref{eq:fulldisp-rel}).
This assertion might seem obvious but it is not~: A $Q \overline{Q}$ cut on the
right--hand fermion loop in Fig.~\ref{fig:3gluon}(a) does not provide,
by itself, a pure real contribution. Only
when both $Q \overline{Q}$ cuts, on the left--hand and right--hand fermion
loops of Fig.~\ref{fig:3gluon}(a), are added we get 
$\mbox{Im} \Pi_{Q \overline{Q}}$.
\par
Let us go back then to Eq.~(\ref{eq:fulldisp-rel}). All the difficulty
with the phenomenological application of the sum rules is now the fact that
the contribution from the three--gluon cut is contained in both sides of
the equality. Thus we propose an {\it infrared safe}
definition of the moments by the trivial subtraction~:
\begin{eqnarray}
\widehat{\Pi}_{ Q\overline{Q}}(q^2)& \doteq & \, \widehat{\Pi} (q^2)-
\Frac{q^2}{\pi}\int^{\infty}_{0} \Frac{ds}{s}\,
\frac{\mbox{Im}\,\Pi_{3g}(s)}{s-q^2-i\epsilon} \; = \; 
\Frac{q^2}{\pi}\int^{\infty}_{4M^2} \Frac{ds}{s}\,
\frac{\mbox{Im}\,\Pi_{ Q\overline{Q}}(s)}{s-q^2-i\epsilon}
\; \; ,
\label{eq:safe_def_pol}\\[5mm]
\widetilde{\cal M}_n & \doteq & 
{\cal M}_n-
\frac{1}{\pi}\int^{\infty}_{0}ds\,
\frac{\mbox{Im}\,\Pi_{3g}(s)}{s^{n+1}}\, \; \; .
\label{eq:safe_def_moments}
\end{eqnarray}
Of course Eqs.~(\ref{eq:safe_def_pol}) and ~(\ref{eq:safe_def_moments})
are meaningless unless we give a precise prescription about how to subtract the
contribution of the massless cuts represented by $\mbox{Im}\,\Pi_{3g}$.
Our previous discussion gives us the tool to proceed.
Once the full $\Oa$ $\Pi(s)$ is calculated we can 
extract the imaginary part starting at $s=0$ (which should go
with a $\theta(s)$ function) for any value of $s$. 
It is clear that the $\theta(s)$ and $\theta(s-4M^2)$ terms in the imaginary
part of the vacuum polarization function correspond to three--gluon
massless and
to $Q\overline{Q}$ cut graphs, respectively, and $\mbox{Im}\,\Pi_{3g}$
and $\mbox{Im}\,\Pi_{ Q\overline{Q}}$ are easy to distinguish, as 
Eq.~(\ref{eq:unita}) prevents the appearance of mixed
\mbox{$\theta(s)\cdot\theta(s-4M^2)$} terms.
Therefore we 
identify 
$\mbox{Im}\,\Pi_{3g}$ and we now plug it in the dispersion
integral of the right--hand side of Eq.~(\ref{eq:safe_def_moments}) 
and perform such
integration. Divergences contained in both this integral and 
${\cal M}_n$ as $q^2\to 0$ will cancel with each other if the same
infrared regularization is employed in the two quantities. 
The intuitive choice would be a low-energy cutoff $s_0 > 0$, and 
Eq.~(\ref{eq:safe_def_moments}) would be more precisely written as:
\begin{equation}
\widetilde{\cal M}_n  \; \equiv \;  \lim_{s_0\to 0^+}\left[
\frac{1}{n!}\left(\frac{d}{dq^2}\right)^n\Pi(q^2)\arrowvert_{q^2 = -s_0} \, 
- \,
\frac{1}{\pi}\int^{\infty}_{0} \Frac{ds}{s} \,
\frac{\mbox{Im}\,\Pi_{3g}(s)}{(s+s_0)^{n}}\right]\; \; ,
\label{eq:safe_moments_reg}
\end{equation}   
where a vanishing therm in the $s_0 \rightarrow 0^+$ limit has been 
omitted.
\par
The evaluation of
the ${\cal M}_n$ moments at $q^2=0 < 4 M^2$ made sense because, up to 
${\cal O}(\alpha_s^2)$, this point is unphysical and the moments
are well defined through an analytic continuation from the high--energy
region. However note that the absorptive three--gluon contribution starts at
$q^2=0$ where perturbative QCD is unreliable. This introduces a further
new difficulty in evaluating ${\cal M}_n$ moments
at $q^2 = 0$, as we reach the physical non--perturbative region. Our
definition of the moments, $\widetilde{\cal M}_n$ in 
Eq.~(\ref{eq:safe_def_moments}), skips this problem by fully 
eliminating the massless terms and, therefore, the final heavy quark sum 
rule will
only involve physics at $q^2 > 4 M^2$.

\begin{figure}[tb]
\begin{center}
\hspace*{-0.5cm}
\includegraphics[angle=0,width=0.85\textwidth]{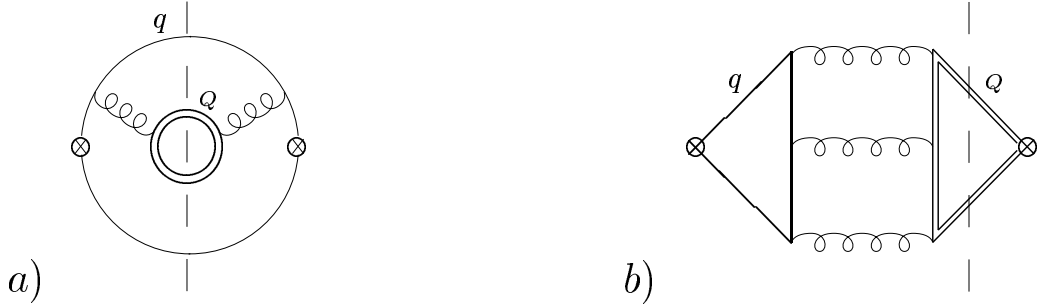}
\end{center}
\caption[]{\label{fig:5gluon} \it Examples of perturbative non--heavy
quark current correlators at ${\cal O} (\alpha_s^2)$ (a) and 
$\Oa$ (b) that contribute to the production of $Q \overline{Q}$ states.}
\end{figure}

The method discussed in this Section could be extended to more general
two--point correlators involving intermediate $Q \overline{Q}$ states in 
their perturbative expansion. Indeed the correlator of two light quark
vector currents has ${\cal O} (\alpha_s^2)$ contributions with an internal
loop of heavy quarks (Fig.~\ref{fig:5gluon} (a)) and, similarly,
the asymmetric correlator of a 
heavy and a light vector quark currents is no longer vanishing at $\Oa$
(Fig.~\ref{fig:5gluon} (b)).
The absorptive part coming from the $Q \overline{Q}$ cuts in the previous 
examples contribute to the phenomenological input 
$\sigma ( e^+ e^- \rightarrow Q \overline{Q})$ in the usual sum rule
analysis. Correspondingly they should be accounted for in the theoretical
side. In short, the production of $Q \overline{Q}$ states concerns not only
the correlator of a couple of heavy quark currents and, for a more rigorous
use of the sum rules method, this imbalance should be taken into account
and properly fixed.
A two--point function built from the sum of the electromagnetic 
currents associated to each quark flavour could be used in a generalized
version of Eq.~(\ref{eq:vacumm})~:
\begin{equation}
\Pi_{\mu \nu}^{G}(p) \; = \; i \, \int \, d^4x \, e^{ipx} \,
\sum_{q,q'} \, e_q \, e_{q'} \, \langle \, 0 \, | \, 
T \, \left( \, \overline{q}(x) \, 
\gamma_{\mu} \, 
q(x) \, \right) \, ( \,  \overline{q'}(0) \, \gamma_{\nu} \, q'(0) \,)
 \, | \, 0 \,
\rangle \; \; \, , 
\label{eq:generalo}
\end{equation}
where now $q$ and $q'$ stand for heavy or light quarks indistinctly,
with electric charges $e_q$ and $e_{q'}$, and
$q = q'$ is also allowed.
As the different absorptive cuts
contribute additively to $\mbox{Im} \Pi(p^2)$, the unwanted light quark and
gluon
$q \overline{q}$, 
$q \overline{q} g$, $ggg$, \ldots cuts could be identified and, through 
the dispersive technique, subtracted from the full
$\Pi(p^2)$ result. Consequently we are left with every possible $Q \overline{Q}$
intermediate state arising from vector current production. Notwithstanding,
the feasibility of this procedure from a technical point of view appears
rather cumbersome and, at present, the experimental accuracy in the 
measurement of $\sigma(e^+e^- \rightarrow Q \overline{Q})$ cannot 
accommodate the corrections just discussed.

\section{Discussion on QCD sum rules applications}
\begin{figure}[tb]
\begin{center}
\hspace*{-0.5cm}
\includegraphics[angle=0,width=0.3\textwidth]{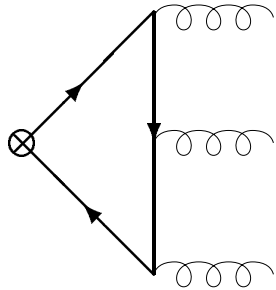}
\end{center}
\caption[]{\label{fig:4gluon} \it Feynman diagram for the production of
three gluons at ${\cal O} (\alpha_s^{3})$.}
\end{figure}

\hspace*{0.5cm}
The general rule given above is valid for all orders of perturbation
theory, but it strongly relies in our ability to extract the
massless absorptive part from the full result of $\Pi(q^2)$
calculated at a definite order. Beyond ${\cal O}(\alpha_s^2)$
complete analytical results for the heavy quark correlator would be
cumbersome 
and only numerical approaches may be at hand. In this
sense, it would be convenient to have a method to
calculate $\mbox{Im}\,\Pi_{ Q\overline{Q}}$ only based  
on Feynman graphs. We have already
sketched such a method in the discussion following
Eq.~(\ref{eq:piq2})~:  we just need to sum up all the massless cut
graphs to get $\mbox{Im}\,\Pi_{3g}$, and then proceed with the dispersion
integration that gives the associated dispersive part \cite{perhaps}.
For example, at 
$\Oa$, the only massless absorptive part comes from the three--gluon
cut in the diagram of Fig.~\ref{fig:3gluon}(a); let us call 
${\cal M}_{3g}^{\mu}$
the amplitude producing three gluons from the heavy quark current at
lowest order (i.e. through the quark triangle loop in
Fig.~\ref{fig:4gluon}). The massless contribution to the absorptive part
of the
correlator is then:
\begin{equation}
\mbox{Im}\,\Pi_{3g} (s) = -\frac{1}{6s}
\int dR_{3g} \,\,{\cal M}_{3g}^{\mu} \cdot {\cal M}_{3g\,\mu}^* 
\; \; ,
\label{eq:alpha3_Img}
\end{equation}  
with the three--gluon phase space integral defined as 
\begin{equation}
\int  dR_{3g} \equiv \frac{1}{3!}\frac{1}{(2\pi)^5}\frac{\pi^2}{4s}
\int_0^s ds_1 \int_0^{s-s_1} ds_2
\, \; \, ,
\label{eq:3gluon_space}
\end{equation}
in terms of the invariants $s_1\equiv (k_1+k_2)^2=(q-k_3)^2$ and
$s_2\equiv (k_2+k_3)^2=(q-k_1)^2$, and $k_i$ being the momenta of
the gluons. The real part would be obtained by integrating 
Eq.~(\ref{eq:alpha3_Img})~:
\begin{equation}
\Frac{s_0}{\pi}\int^{\infty}_{0} \Frac{ds}{s} \,
\frac{\mbox{Im}\,\Pi_{3g}(s)}{s+s_0} =
\Frac{-s_0}{288(2\pi)^4}\int^{\infty}_{0}
\frac{ds}{s^3(s+s_0)}\int_0^s ds_1 \int_0^{s-s_1}ds_2
\,\,{\cal M}_{3g}^{\mu} \cdot {\cal M}_{3g\,\mu}^*
\,,
\label{eq:alpha3_Reg}
\end{equation}
which, in principle, could be performed also numerically.  
The $n$th-derivative of relation (\ref{eq:alpha3_Reg}) respect 
to $s_0$, in the limit $s_0\to 0^+$, would give the infrared divergent
contribution that should be subtracted from the full moments,
as dictated by Eq.~(\ref{eq:safe_moments_reg}).  
\par
This evaluation solves the extraction procedure on the theoretical
side. However the use of moments on heavy quark sum rules involves 
global quark--hadron duality which translates into the supposed
equivalence between the theoretical
evaluation of the moments and the phenomenological input. 
The implementation of the latter is not a trivial task. 
The total experimental cross section 
$\sigma (e^+ e^- \rightarrow hadrons)$ can be split into two disjoint
quantities~: the cross section for producing hadrons with
Q--flavoured states, and the production of hadrons with no Q--flavoured
components. If the experimental set up was accurate
enough to classify events into one of these two clusters, the first
class would be the required ingredient for the phenomenological part
of the heavy quark sum rule. However this separation, implemented in 
the theoretical
side within perturbative QCD, is rather involved. Up to $\Oaa$ there
has not been any doubt, in the literature, that contributions to this
side arise wholly from $Q \overline{Q}$ cuts in the heavy quark 
correlator $\Pi_{Q \overline{Q}}$. 
The physical picture behind this assertion relies in the assumption
of factorization between hard and soft regions in the quark production
process and subsequent hadronization. The hard region described
with perturbative QCD entails the production of the pair of heavy
quarks, and the soft part of the interaction is responsible for the observed
final hadron content. Although possible, annihilation of the partonic
state $Q \overline{Q}$ due to the later interaction is very unlikely,
as jets arising from the short distance interaction fly apart before
long--distance effects become essential. Consequently, each jet
hadronizes to a content of Q--flavoured states with unit probability.
As local duality is implicitly
invoked, this picture is assumed to hold at sufficient high energies; hence
perturbative corrections to the hard part are successively included
through the heavy quark currents correlator.

The same physical picture applied to the three-gluon cut does not allow us
to conclude whether this piece should be included or not in the
theoretical side of the sum rules.  
Confinement tell us that this intermediate state hadronizes completely
into hadrons with a 
content of light and/or heavy quarks indistinctly.
It is conspicuous that if we could disentangle the heavy quark hadronization,
$3g \rightarrow Q \overline{Q}$, we should include only this piece into
the sum rule. Then the singularity at $q^2 = 0$ would disappear because 
heavy quarks are produced starting at $q^2 = 4M^2$. However there is no
way to sort out light and heavy quark production off three gluons and,
therefore, if we extract
this contribution from the heavy quark sum rules we are introducing
an incertitude in the procedure because we make sure that there is no 
light quark hadronization but we miss the heavy quark production. 
It is easy to see that the induced error is small, due to the fact that
three gluons hadronize mostly to light hadrons. On one side, in the very
high energy region and following perturbative QCD with $N_F = 4$,
we have only a $1/4 \, = \, 25 \, \%$ probability of finding
a specified pair of heavy quarks produced. And this is a generous upper limit
because when we go down in energy, phase space restrictions severely reduce
the counting of heavy quarks. Hence we estimate that excluding
the three--gluon cut we introduce a tiny very few percent error in the
sum rules procedure.

\section{Conclusions}
\hspace*{0.5cm}
We have shown that
rigorous and straightforward results of the general theory of 
singularities of perturbation theory amplitudes provide 
all--important tools to extract the unwanted $\Oa$ three--gluon massless
cut pointed out by Groote and Pivovarov from the vector current
correlator of heavy quarks.
We conclude that the appropriate procedure to obtain information 
about the heavy quark parameters should make use of the infrared safe
corrected 
moments, defined in Eq.~(\ref{eq:safe_moments_reg}), that now indeed
satisfy the modified sum rule~:
\begin{equation}
\widetilde{\cal M}_n \; = \; \frac{1}{\pi}\int^{\infty}_{4 M^2}ds\,
\frac{\mbox{Im}\,\Pi_{Q\overline{Q}}(s)}{s^{n+1}} \; \; ,
\label{eq:finali}
\end{equation}
where the right--hand side can be extracted from the heavy quark
production cross section $\sigma(e^+e^- \rightarrow Q \overline{Q})$.
\par
Finally we have pointed out that, starting already at 
${\cal O} (\alpha_s^2)$, the use of sum rules associated to heavy
vector current correlators shows and imbalance between the phenomenological
input in the dispersion relation and the perturbative two--point
function. This is due to the fact that the cross section of production of
$Q \overline{Q}$ heavy quarks is contained not only in the correlator of 
two heavy quark currents but in those involving light quarks too. We have
indicated how to improve the application of sum rules
by constructing a generalized correlator of vector currents in 
Eq.~(\ref{eq:generalo}) and, afterwards, extracting all the perturbative
information not related with the production of heavy quarks, as we did
in detail for the three--gluon cut.
\vspace*{1cm} \\
\noindent
{\large \bf Acknowledgements}\par
\vspace{0.2cm}
\noindent 
We wish to thank A.~Pich
for calling our attention on this problem.
We also thank G.~Amor\'os, M.~Eidem\"uller and A.~Pich for relevant 
discussions on the
topic of this paper and for reading the manuscript. 
The work of P.~D. Ruiz-Femen\'\i a has been partially supported by a FPU
scholarship of the Spanish {\it Ministerio de Educaci\'on y Cultura}.
This work has been supported in part by TMR, EC Contract No. 
ERB FMRX-CT98-0169, by MCYT (Spain) under grant FPA2001-3031, and by
ERDF funds from the European Commission.


\begin{thebibliography}{99}

\bibitem{pg1} S.~Groote and A.~A.~Pivovarov,
	      JETP Lett.\  {\bf 75} (2002) 221
              [Pisma Zh.\ Eksp.\ Teor.\ Fiz.\  {\bf 75} (2002) 267]
              [arXiv:hep-ph/0103047].
	       
\bibitem{dera} T.~Appelquist and H.~Georgi, 
               Phys. \ Rev. \ {\bf D8} (1973) 4000; \\
	       A.~Zee,
	       Phys. \ Rev. \ {\bf D8} (1973) 4038.

\bibitem{rev} M.~A.~Shifman, A.~I.~Vainshtein and V.~I.~Zakharov,
              Nucl.\ Phys.\ B {\bf 147} (1979) 385; \\
	      M.~A.~Shifman, A.~I.~Vainshtein and V.~I.~Zakharov,
              Nucl.\ Phys.\ B {\bf 147} (1979) 448; \\
	      L.~J.~Reinders, H.~Rubinstein and S.~Yazaki,
              Phys.\ Rept.\  {\bf 127} (1985) 1; \\
              M.~Jamin and A.~Pich,
              Nucl.\ Phys.\ B {\bf 507} (1997) 334
              [hep-ph/9702276]; \\
	      A.~H.~Hoang,
	      Phys.\ Rev.\ D {\bf 59} (1999) 014039
              [hep-ph/9803454]; \\
	      M.~Beneke and A.~Signer,
	      Phys.\ Lett.\ B {\bf 471} (1999) 233
              [hep-ph/9906475]; \\
              M.~Eidemuller and M.~Jamin,
	      Phys.\ Lett.\ B {\bf 498} (2001) 203
	      [hep-ph/0010334].

\bibitem{cher} K.~G.~Chetyrkin, J.~H.~Kuhn and M.~Steinhauser,
		Nucl.\ Phys.\ B {\bf 482} (1996) 213
		[hep-ph/9606230]; \\
		K.~G.~Chetyrkin, R.~Harlander, J.~H.~Kuhn and M.~Steinhauser,
		Nucl.\ Phys.\ B {\bf 503} (1997) 339
		[hep-ph/9704222].

\bibitem{pg2} S.~Groote and A.~A.~Pivovarov,
	      Eur.\ Phys.\ J.\ C {\bf 21} (2001) 133
              [arXiv:hep-ph/0103313].

\bibitem{differ}D.~Z.~Freedman, K.~Johnson and J.~I.~Latorre,
		Nucl.\ Phys.\ B {\bf 371} (1992) 353; \\
	         D.~Z.~Freedman, G.~Grignani, K.~Johnson and N.~Rius,
	         Annals Phys.\  {\bf 218} (1992) 75
		[hep-th/9204004].

\bibitem{oldies} L.~D.~Landau, Nucl. \ Phys. \ {\bf 13} (1959) 181; \\
                 J.~C.~Taylor, Phys. \ Rev. \ {\bf 117} (1960) 261; \\
		 R.~E.~Cutkosky, J. \ Math. \  Phys. \ {\bf 1} (1960) 429; \\
		 R.~E.~Cutkosky, Rev. \ Mod. \ Phys. \ {\bf 33} (1961) 448. 

\bibitem{perhaps} J.~Portol\'es and P.~D.~Ruiz--Femen\'\i a, work in progress.

\end{thebibliography}
\end{document}